\documentclass[aps,pre,twocolumn,showpacs,preprintnumbers,a4paper]{revtex4} 

\usepackage{graphics}
\usepackage{epsf}
\usepackage{epsfig}

\begin{document} 

\flushbottom
\def\bottomfraction{0.5}

\title{Scaling behavior of the conserved transfer threshold process}

\author{S. L\"ubeck}
\email{sven@thp.uni.duisburg.de}
\affiliation{Weizmann Institute of Science, 
Department of Physics of Complex Systems, 
76100 Rehovot, Israel,\\
Theoretische Tieftemperaturphysik, 
Gerhard-Mercator-Universit\"at,
47048 Duisburg, Germany 
}

\date{June 20, 2002}

\begin{abstract}
We analyze numerically the critical behavior of an absorbing
phase transition in the
conserved transfer threshold process.
We determined the steady state scaling behavior of the order parameter
as a function of both, the control parameter and an 
external field, conjugated to the order parameter.
The external field is realized as a spontaneous creation
of active particles which drives the system away from
criticality.
The obtained results yields that the conserved transfers
threshold process belongs to the universality class of
absorbing phase transitions in a conserved field.
\end{abstract}



\preprint{accepted for publication in {\it Physical Review E} 2002}

\maketitle

\section{Introduction}

The scaling behavior of directed percolation is recognized 
as the paradigm of the critical behavior
of several non-equilibrium systems which exhibits a 
continuous phase transition from an active state
to an absorbing non-active state 
(see for instance~\cite{HINRICHSEN_1}).
The widespread occurrence of such systems in
physics, biology, as well as catalytic chemical 
reactions is reflected by the well known universality
hypothesis of Janssen and Grassberger that models which 
exhibit a continuous phase transition to a single
absorbing state generally belong to the universality class
of directed percolation~\cite{JANSSEN_1,GRASSBERGER_2}.
Introducing additional symmetries the critical
behavior differs from directed percolation.
In particular particle conservation leads to
the different universality class of absorbing phase transitions
with a conserved field as pointed out in~\cite{ROSSI_1}.
In that work the authors introduced two models,
the conserved lattice gas (CLG) and the 
conserved threshold transfer process (CTTP).
The latter one is a conserved modification
of the threshold transfer process introduced
in~\cite{MENDES_1}.
Both models display a continuous phase transition
from an active to an inactive phase
and are believed to belong to the same 
universality class~\cite{ROSSI_1}.
The steady-state scaling behavior of the CLG model
was investigated recently.
The order parameter and its fluctuations were
numerically examined in~\cite{LUEB_19}.
The scaling behavior in an external field conjugated 
to the order parameter
was considered in~\cite{LUEB_22}.
Furthermore a modified CLG model was introduced
which allows to determine analytically the steady-state 
mean-field scaling behavior of the universality 
class~\cite{LUEB_20,LUEB_25}.

On the other hand the scaling behavior of
the CTTP was investigated in low dimensional
($D=1,2$) systems only~\cite{ROSSI_1,DICKMAN_3} 
and no external field was applied.
Therefore we consider in this work
the CTTP with and without an external field
in various dimensions~($D=1,2,3,4,5,6$)
and determine a set of critical exponents.
All obtained results coincides with those
of the CLG model, strongly supporting the universality
hypothesis of~\cite{ROSSI_1}.

\section{Model and scaling behavior}
\label{sec:model}

In this work we consider the CTTP on simple
cubic lattices of linear size~$L$ 
in various dimensions with~$N$ particles.
The lattice sites may be empty, occupied 
by one particle, or occupied by two particles.
Empty and single occupied sites are considered
as non-active whereas double occupied lattice sites are considered as
active.
In the latter case one tries to transfer both particles of 
each active site to randomly chosen empty or single occupied
nearest neighbor sites.
If no active sites exist the system is trapped forever
in a certain configuration, a so-called absorbing state.

In the following we denote the densities of active sites with
$\rho_{\text a}$
and the density of particles on the lattice as
$\rho=N/L^D$, which is considered as the 
control parameter of the absorbing phase
transition.
The density of active sites $\rho_{\text a}$ is the
order parameter of the absorbing phase transition, i.e., it
vanishes at the critical
density $\rho_{\text c}$ according to
\begin{equation}
\rho_{\text a} \sim \delta\rho^{\beta},
\label{eq:order_par}
\end{equation}
with the reduced control parameter 
$\delta\rho=\rho/\rho_{\text c}-1$ and the 
order parameter exponent~$\beta$.

Similar to equilibrium phase transitions it is 
possible in the case of absorbing phase transitions
to apply an external field~$h$ which is
conjugated to the order parameter 
(see for instance~\cite{HINRICHSEN_1}).
As usually for continuous phase transitions
the conjugated field has to destroy the disordered
phase and the associated linear response function
$\partial \rho_{\text a}/ \partial h$
has to diverge at the critical point ($\delta\rho=0$, $h=0$).
In the case of an absorbing phase transition the external 
field acts as
a spontaneous creation of active particles, i.e.,
the external field destroys the absorbing state
and thus the phase transition itself.
But considering absorbing phase transitions with 
particle conversation one has 
to take care that the external field does not
change the particle number.
A possible realization of the external field 
was developed in~\cite{LUEB_22} where the
external field triggers movements of inactive
particles which may be activated in this way.
The external field~$h$ is another
relevant scaling field and for sufficiently
small values of~$h$ the order parameter scales
as
\begin{equation}
\rho_{\rm{a}}(\delta\rho, h) \; \sim \; 
\lambda\, \, {\tilde r}
(\delta \rho \; \lambda^{-1/\beta}, h \; \lambda^{-\sigma/\beta})
\label{eq:scal_ansatz}
\end{equation}
with the critical field exponent $\sigma$
and the scaling function~${\tilde r}$.
Choosing $\delta\rho \, \lambda^{-1/\beta}=1$ one recovers
Eq.\,(\ref{eq:order_par})
whereas $h \, \lambda^{-\sigma / \beta}=1$ leads 
at the critical density to 
\begin{equation}
\rho_{\rm a} \sim h^{\beta/\sigma}.
\label{eq:order_par_field}
\end{equation}

In our simulations we start with randomly distributed
particles.
All active sites are listed and this list is 
updated in a randomly chosen sequence.
In the case that an external field is applied
the active particle creation is performed
after each update step in order to mimic the external field.
After a certain relaxation time the system reaches a steady
state where the density of active sites at update step
$t$ fluctuates around the average value 
$\langle \rho_{\text a}(\delta\rho,h,t)\rangle$
which is interpreted as the order parameter $\rho_{\text a}(\delta\rho,h)$
(see for instance Figs.\,1 of~\cite{LUEB_19,LUEB_22}).


Additionally to the order parameter we consider its 
fluctuations 
\begin{equation}
\Delta \rho_{\text a}(\delta\rho, h) 
\; = \; L^D \, \left [ \langle \rho_{\text a}(\delta\rho, h,t)^2 \rangle
\, - \, \langle \rho_{\text a}(\delta\rho,h,t)\rangle^2 \right ].
\label{eq:def_fluc}
\end{equation}
Approaching the transition point the fluctuations diverge 
for zero-field according to
\begin{equation}
\Delta \rho_{\text a}(\delta\rho, h=0) 
\; \sim \; \delta\rho^{-\gamma^{\prime}}.
\label{eq:fluc_crit}
\end{equation}
The fluctuation exponent $\gamma^{\prime}$
fulfills the scaling relation~\cite{JENSEN_3}
\begin{equation}
\gamma^{\prime} \; = \; \nu_{\scriptscriptstyle \perp} D \, - \, 
2 \, \beta,
\label{eq:scal_rel_gamma_prime}
\end{equation}
where the exponent $\nu_{\scriptscriptstyle \perp}$
describes how the spatial correlation length diverges 
at the transition point.
In the critical regime we assume that the fluctuations
obey the scaling ansatz
\begin{equation}
\Delta \rho_{\text a}(\delta\rho, h) 
\; = \; 
\lambda^{\gamma^{\prime}}\, \, {\tilde d}
(\delta \rho \; \lambda, h \, \lambda^{\sigma}) .
\label{eq:fluc_scal_ansatz}
\end{equation}
Setting $\delta \rho \, \lambda=1$ one recovers 
Eq.\,(\ref{eq:fluc_crit}) for $h=0$.

Analogous to equilibrium phase transitions the susceptibility
is defined as the derivative of the order parameter with respect to the
conjugated field
\begin{eqnarray}
\label{eq:def_suscept}
\chi(\delta\rho,h) &  =  & \frac{\partial\hphantom{h}}{\partial h} \,
\rho_{\text a}(\delta\rho, h) \nonumber \\
& = &
\lambda^{1-\sigma/\beta}\, \, {\tilde c}
(\delta \rho \; \lambda^{-1/\beta}, h \; \lambda^{-\sigma/\beta}).
\end{eqnarray}
Setting $\delta\rho\, \lambda^{-1/\beta}=1$ one gets that the 
susceptibility diverges for zero-field as required according to
\begin{equation}
\left . \frac{\partial \rho_{\text a}}{\partial h_{\phantom a}} 
\right |_{h \to 0}
\; \sim \;
\delta\rho^{-\gamma} .
\label{eq:suscept_gamma}
\end{equation}
Furthermore, one yields the scaling relation
\begin{equation}
\gamma \; = \; \sigma \, - \,  \beta
\label{eq:widom}
\end{equation}
which corresponds to the well known Widom equation  
of equilibrium phase transitions.
Using this scaling relation one can
calculate the value of the 
susceptibility exponent~$\gamma$ from the obtained values
of $\beta$ and $\sigma$.
Notice that in contrast to the scaling behavior of 
equilibrium phase transitions
the non-equilibrium absorbing phase transition 
is characterized by $\gamma \neq \gamma^{\prime}$.

\section{Below the upper critical dimension}
\label{sec:d1_d2_d3}

\begin{figure}[b]
  \includegraphics[width=8.0cm,angle=0]{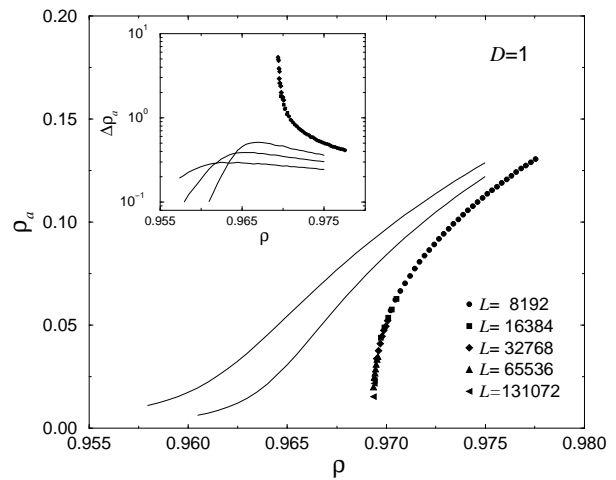}
  \caption{
    The order parameter~$\rho_{\text a}$ as a function of the
    particle density for zero-field (symbols) and for 
    $h=0.0001$ and $h=0.00005$ (lines).
    The inset displays the order parameter 
    fluctuations~$\Delta \rho_{\text a}$ for zero field (symbols)
    and for $h=0.0002$, $h=0.0001$ and $h=0.00005$ (lines).
   }
  \label{fig:rho_a_1d} 
\end{figure}

At the beginning of our analysis we consider the
scaling behavior of the order parameter 
for $D=1,2,3$.
System sizes up to $L=131072$ for $D=1$, $L=2048$ for 
$D=2$, and $L=256$ for ($D=3$) are considered.
In each cases we start the simulation
with randomly distributed particles.
After a certain transient regime the system reaches
a steady state where the density of active 
particles fluctuates around an average 
value which is interpreted as the order parameter.
In the steady state up to $2\, 10^8$ update
steps for $D=1$ and $2\, 10^6$ for $D=2,3$
are performed to measure the average density
of active sites.
For zero-field this procedure is repeated 
for at least 10 different initial configurations
in order to get an accurate 
estimation of the order parameter close to the
critical point ($\rho=\rho_{\text c}$, $h=0$).

In Fig.\,\ref{fig:rho_a_1d} we present the data
of the one-dimensional order parameter at zero-field.
Approaching the transition point the corresponding 
correlation length increases and the system tends to 
the absorbing state if the correlation length is of the 
order of the system size.
Instead of a finite-size scaling analysis 
(see for instance~\cite{ROSSI_1,DICKMAN_3,DICKMAN_2})
we take care of these finite-size effects in the way that
we increase the system size before these finite-size effects 
occur and use only data from simulations that have not reached
the absorbing state.

\begin{figure}[t]
  \includegraphics[width=8.0cm,angle=0]{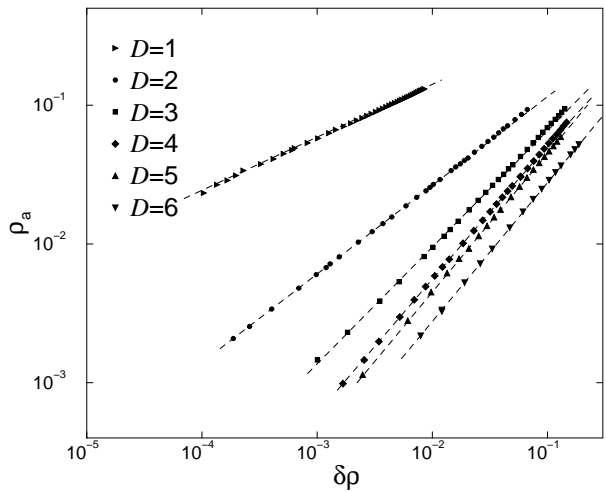}
  \caption{
    The order parameter~$\rho_{\text a}$ as a function of the
    reduced particle density $\delta\rho$ at zero-field
    for various dimensions~$D$.
    The dashed line corresponds to a power-law behavior
    according to Eq.\,(\protect\ref{eq:order_par}) for $D\neq D_{\text c}$.
    For $D=6$ the data are shifted horizontally by a factor 1.5
    in order to avoid an overlap. 
    In the case of the four-dimensional model
    the dashed line corresponds to 
    Eq.\,(\protect\ref{eq:order_par_dc_zero_field})
    with ${\text B}= 0.15$.
   }
  \label{fig:rho_a_all_d} 
\end{figure}

\begin{figure}[b]
  \includegraphics[width=8.0cm,angle=0]{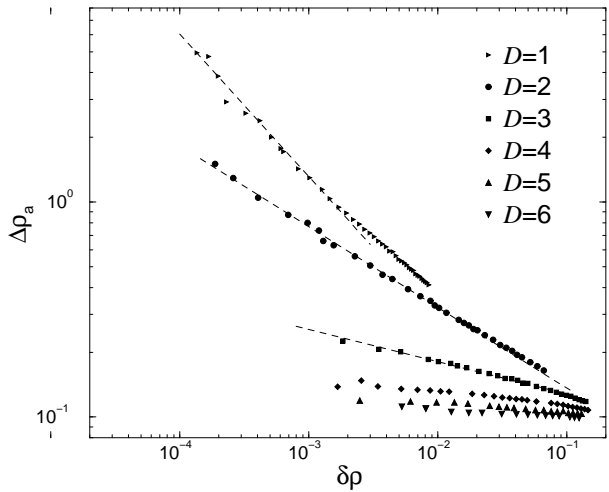}
  \caption{
    The order parameter fluctuations~$\Delta\rho_{\text a}$
    as a function of the reduced particle density $\delta\rho$
    at zero-field for various dimensions~$D$.
    The dashed line corresponds to a power-law divergence
    [Eq.\,(\protect\ref{eq:fluc_crit})].
    For $D\ge D_{\text c}$ the fluctuations are maximal at
    the transition point but finite.
   }
  \label{fig:rho_a_fluc_all_d}
\end{figure}

Decreasing the particle density
the order parameter decreases and
vanishes at the transition point.
To determine the critical indices
one varies the critical density~$\rho_{\text c}$ until
one obtains asymptotically a straight line in a log-log plot.
The exponent is then obtained by a regression
analysis.
The values of the order parameter as a function of the
reduced particle density $\delta \rho$
are plotted in Fig.\,\ref{fig:rho_a_all_d}.
In all cases the asymptotic behavior ($\delta\rho\to 0$) 
of the order parameter obeys Eq.\,(\ref{eq:order_par}).
For $D=1$ we get $\rho_{\text c}=0.96929 \pm 0.00003$
and $\beta=0.382 \pm 0.019$. 
The latter value is smaller than the value 
$\beta=0.412$ 
estimated from significantly smaller system sizes 
($L\leq 5000$)~\cite{DICKMAN_3}.
Furthermore our value differs from $\beta=0.42 \pm 0.02$ obtained
from simulations of the one-dimensional fix-energy Manna
sandpile model~\cite{DICKMAN_2} that is expected to belong 
to the same universality class.

In the two dimensional case we obtain $\rho_{\text c}=0.69392\pm 0.00001$
and $\beta=0.639\pm 0.009$.
Again the order parameter exponent differs slightly
from the previously reported result $\beta=0.656$ 
obtained from simulations of small lattice 
sizes ($L\leq 160$)~\cite{DICKMAN_3}. 
But our value agrees with the estimate of the corresponding
two-dimensional Manna sandpile 
model $\beta=0.64\pm 0.01$~\cite{VESPIGNANI_4}.

The estimates of the three dimensional model 
are $\rho_{\text c}=0.60489\pm 0.00002$
and $\beta=0.840\pm 0.012$.
All obtained critical exponents are listed in 
Table~\ref{table:critical_indicees}.

In Fig.\,\ref{fig:rho_a_fluc_all_d} we present the
order parameter fluctuations as a function of the
control parameter at zero field.
We observe for $D< D_{\text c}$ a power-law behavior
according to Eq.\,(\ref{eq:fluc_crit}).
Using a regression analysis we get the estimates
$\gamma^{\prime}=0.662 \pm 0.071$ for $D=1$,
$\gamma^{\prime}=0.381 \pm 0.013$ for $D=2$, and
$\gamma^{\prime}=0.187 \pm 0.030$ for $D=3$.

\begin{figure}[t]
  \includegraphics[width=8.0cm,angle=0]{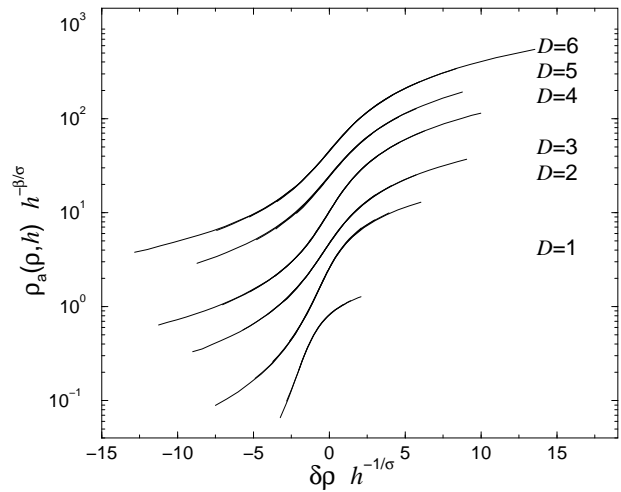}
  \caption{
    The scaling plot of the order parameter fluctuations~$\Delta\rho_{\text a}$ 
    for various dimensions.
    For $D>1$ the curves are shifted vertically in order to avoid overlaps.
    In the case of the four-dimensional model
    $\rho_{\text a} \, h^{-1/2} \,| \ln{h}|^{-\Sigma}$ is plotted 
    vs.~$\delta\rho\, h^{-1/2} |\ln{h}|^{b-s/2}$
    with $\Sigma=0.28$ and $b-s/2=-0.12$ (see text).
   }
  \label{fig:rho_a_field_scal_all_d}
\end{figure}

In the following we analyze the order parameter as a function
of the control parameter $\delta\rho$ for different fields
from $h=10^{-5}$ up to $2\,10^{-4}$.
The applied field results in 
a smoothing of the zero-field curve, i.e., the order parameter
increases smoothly with the control parameter for $h>0$
(see Fig.\,\ref{fig:rho_a_1d}).
According to the scaling ansatz of the order parameter
[Eq.\,(\ref{eq:scal_ansatz})] we choose $h \, \lambda^{-\sigma/\beta}=1$
and get the scaling form
\begin{equation}
\rho_{\text{a}}(\delta\rho, h) \; = \; 
h^{\beta/\sigma}\, \, {\tilde r}
(\delta \rho \; h^{-1/\sigma}, 1).
\label{eq:scal_plot_order}
\end{equation}
Thus one varies the exponent $\sigma$ until the 
curves for different values of the driving field have to collapse
onto the scaling function ${\tilde r}$ if one plots
$\rho_{\text a} \,h^{-\beta/\sigma}$ 
as a function of $\delta \rho \, h^{-1/\sigma}$.
Convincing results are obtained for 
$\sigma=1.770 \pm 0.058$ ($D=1$),
$\sigma=2.229 \pm 0.032$ ($D=2$), as well as 
$\sigma=2.069 \pm 0.043$ ($D=3$) and 
the corresponding scaling plots are shown in 
Fig.\,\ref{fig:rho_a_field_scal_all_d}.

\begin{figure}[t]
  \includegraphics[width=8.0cm,angle=0]{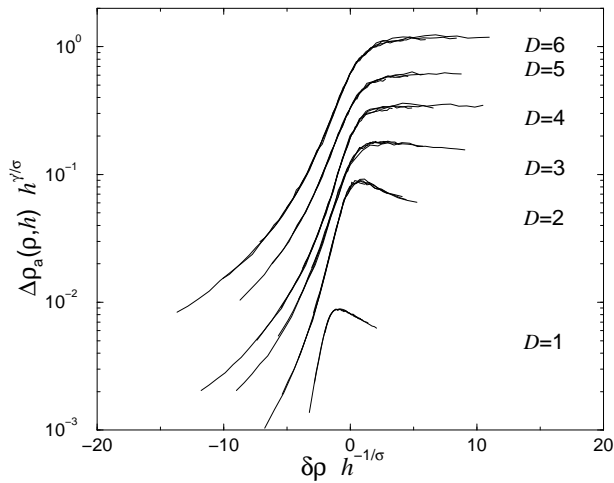}
  \caption{
    The scaling plots of the order parameter 
    fluctuations~$\Delta\rho_{\text a}$ 
    for various dimensions.
    For $D>1$ the curves are shifted vertically in order to avoid overlaps.
    The fluctuations diverges at the critical point for $D<4$
    whereas a jump of the fluctuations is observed in higher dimensions
    at zero-field.
    In the in case of the four-dimensional model
    $\Delta\rho_{\text a}$ is plotted 
    vs.~$\delta\rho\, h^{-1/2} \, |\ln{h}|^{-\eta}$
    with $\eta=0.1$.
   }
  \label{fig:rho_a_fluc_scal_all_d}
\end{figure}

Next we consider the scaling behavior of the order
parameter fluctuations $\Delta\rho_{\text a}$.
The fluctuation data for $D=1$ are shown  
for different values of the external field in the 
inset of Fig.\,\ref{fig:rho_a_1d}.
For finite fields the fluctuations display a peak. 
Approaching the transition point ($h\to 0$) this peak
becomes a divergence signalling the critical point.
In order to analyze the scaling behavior of the fluctuations 
we use the scaling ansatz Eq.\,(\ref{eq:fluc_scal_ansatz}) 
and set $h \, \lambda^{\sigma}=1$ 
\begin{equation}
\Delta \rho_{\text a}(\delta\rho, h) 
\; = \; 
h^{-\gamma^{\prime}/\sigma}\, \, {\tilde d}
(\delta\rho \; h^{-1/\sigma},1).
\label{eq:scal_plot_fluc}
\end{equation}
Using the above determined values of $\rho_{\text c}$,
$\sigma$ and $\gamma^{\prime}$ we get good data
collapses confirming the accuracy of our analysis.
(see Fig.\,\ref{fig:rho_a_fluc_scal_all_d})

Furthermore we determine the susceptibility 
exponent~$\gamma$.
Using the scaling relation Eq.\,(\ref{eq:widom}) one gets
the estimates of the susceptibility
exponents 
$\gamma=1.388 \pm0.063$ ($D=1$),
$\gamma=1.590 \pm0.033$ ($D=2$), and 
$\gamma=1.229 \pm0.045$ ($D=3$).

\section{At the upper critical dimension}
\label{sec:dc_d4}

In the case of the four dimensional model
we considered system sizes from $L=8$ up to
$L=64$.
At least $10^6$ update steps were used to 
reach the steady state close to the transition point
and $2\,10^6$ update steps were performed
to determine the order parameter and its
fluctuations.

\begin{figure}[b]
  \includegraphics[width=8.0cm,angle=0]{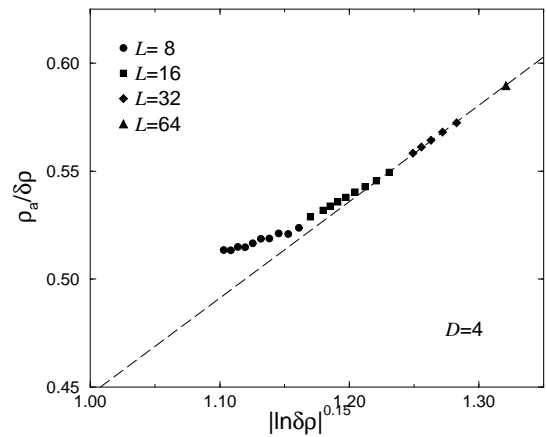}
  \caption{ 
    The density of active sites at the upper critical
    dimension $D_{\text c}=4$.
    The data are rescaled according to 
    Eq.\,(\protect\ref{eq:order_par_dc_zero_field}).
    The assumed asymptotic scaling behavior (dashed line) is obtained
    for ${\text B}=0.15$.
   }
  \label{fig:rho_a_4d_log}
\end{figure}

At the upper critical dimension $D_{\text c}=4$ the scaling behavior
of the CTTP is affected by logarithmic 
corrections similar to the CLG model~\cite{LUEB_19,LUEB_22}. 
As argued in~\cite{LUEB_22} the order parameter
obeys in leading order the scaling ansatz
\begin{eqnarray}
\label{eq:scal_ansatz_dc}
\rho_{\text{a}}(\delta\rho, h) \; = \; & \\ \nonumber 
& \lambda\, |\ln{\lambda}|^l \; {\tilde r}
(\delta \rho \, \lambda^{-1/\beta} |\ln{\lambda}|^b, 
h \, \lambda^{-\sigma/\beta} |\ln{\lambda}|^s), 
\end{eqnarray}
where the exponents $\beta$ and $\sigma$ are given 
by the corresponding mean-field values $\beta=1$ and
$\sigma=2$~\cite{LUEB_25}, respectively.
Thus, for zero field the asymptotic
scaling behavior of the order parameter obeys
\begin{equation}
\rho_{\text{a}}(\delta\rho, h=0) \; \sim \; 
\delta\rho \; | \ln{\delta\rho}|^{\text B}
\label{eq:order_par_dc_zero_field}
\end{equation}
with ${\text B}=b+l$.
In our analysis we plot $\rho_{\text a} / \delta\rho$
as a function of $| \ln{h} |^{\text B}$ and vary the
exponent~${\text B}$ as well as the critical 
density~$\rho_{\text c}$
until one gets asymptotically a straight line. 
The best result is obtained for ${\text B}=0.15$,
$\rho_{\text c}=0.56705\pm 0.00003$
and the corresponding plot is shown 
in Fig.\,\ref{fig:rho_a_4d_log}.
This value of ${\text B}$ differs 
from the corresponding value of the CLG model
${\text B}=0.24$~\cite{LUEB_22}.

Similar to the lower dimensions we consider the 
scaling behavior of the order parameter as a function
of the control parameter for different external fields.
Choosing $h \, \lambda^{-\sigma/\beta} |\ln{\lambda}|^s=1$
the scaling ansatz [Eq.\,(\ref{eq:scal_ansatz_dc})] yields 
in leading order 
\begin{equation}
\rho_{\text{a}}(\delta\rho, h) \; = \; 
h^{1/2} \; | \ln{h}|^{\Sigma} \; 
{\tilde r} (x, 1 ),
\label{eq:order_par_dc_scal}
\end{equation}
where the scaling argument~$x$ 
is given in leading order by
\begin{equation}
x \; = \; 
\delta\rho \; h^{-1/2} \,|\ln{h}|^{b-s/2}
\label{eq:scal_arg_lead_ord}
\end{equation}
with $\Sigma=s/2+l$.
Varying the logarithmic correction exponents
one gets for $\Sigma=0.28$ and $b-s/2=-0.12$ 
a convincing data-collapse, which is shown in
Fig.\,\ref{fig:rho_a_field_scal_all_d}.
Using the values $\Sigma=l+s/2=0.28$ and 
$b-s/2=-0.12$ we get the estimation ${\text B}=b+l=0.16$
which agrees with ${\text B}=0.15$ 
obtained from numerical simulations in zero-field.
On the other hand this values differs 
from the corresponding estimations of the CLG model
${\text B}=0.24$, $\Sigma=0.45$, 
and $b-s/2=-0.17$~\cite{LUEB_22}.

Furthermore we consider how the logarithmic corrections affect
the scaling behavior of the fluctuations at the upper
critical dimension.
As pointed out in~\cite{LUEB_22} the order
parameter fluctuations are expected to obey the
scaling ansatz
\begin{equation}
\Delta \rho_{\text a}(\delta\rho, h) 
\; = \; 
{\tilde d}
(\delta \rho \, h^{-1/2}\,|\ln{h}|^{-\eta} , 1).
\label{eq:fluc_scal_ansatz_dc_02}
\end{equation}
A good data collapse is observed for $\eta=0.10$ 
(see Fig.\,\ref{fig:rho_a_fluc_scal_all_d}) which
differs again from the corresponding
value of the CLG model $\eta=0.39$~\cite{LUEB_22}.

\section{Above the upper critical dimension}
\label{sec:d5}

A modified version of the CTTP with random 
neighbor hopping was recently introduced 
in~\cite{LUEB_25}.
There, unrestricted particle hopping breaks long
range correlations and the scaling behavior is 
characterized by the mean-field values 
$\rho_{\text c}=1/2$, 
$\beta=1$, and $\sigma=2$ which are calculated
analytically.

In our simulations of the five dimensional model
we considered system sizes from $L=8$ up to
$L=32$ whereas system sizes from $L=4$ up to
$L=16$ are used for $D=6$.
At least $2\,10^6$ update steps were used
to reach the steady state and 
$2\,10^6$ update steps were performed
to determine the order parameter and its
fluctuations.
The values of the order parameter
are plotted in Fig.\,\ref{fig:rho_a_all_d}
and the obtained critical densities are
$\rho_{\text c}=0.54864\pm 0.00005$ for 
$D=5$ and $\rho_{\text c}=0.53816\pm 0.00007$
for $D=6$, respectively.
In both dimensions the asymptotic scaling
behavior of the order parameter is in agreement
with the mean-field behavior
$\beta=1$.

The fluctuations of the order parameter $\Delta\rho_{\text a}$
are plotted in Fig.\,\ref{fig:rho_a_fluc_all_d}.
Analogous to the CLG model the fluctuations are
characterized by a jump at the transition point
corresponding to $\gamma^{\prime}=0$~\cite{LUEB_19}.

Above the critical dimension, i.e.~$D\ge 5$, the
scaling behavior of the CTTP is expected to
obey again the scaling ansatzes 
Eqs.\,(\ref{eq:scal_ansatz},\ref{eq:fluc_scal_ansatz})
where the exponents are given by the mean-field values
independently of the particular dimension.
The obtained data collapse of the order parameter 
curves are presented 
in Fig.\,\ref{fig:rho_a_field_scal_all_d}
and confirm the above scenario.

Furthermore we consider the fluctuations above
the upper critical dimension.
According to the mean-field value 
$\gamma^{\prime}=0$~\cite{LUEB_19} we
plot $\Delta \rho_{\text a}$ as a function
of $\delta\rho \, h^{-1/2}$ 
and the obtained data collapses are shown
in Fig.\,\ref{fig:rho_a_fluc_scal_all_d}.

\begin{figure}[t]
  \includegraphics[width=8.0cm,angle=0]{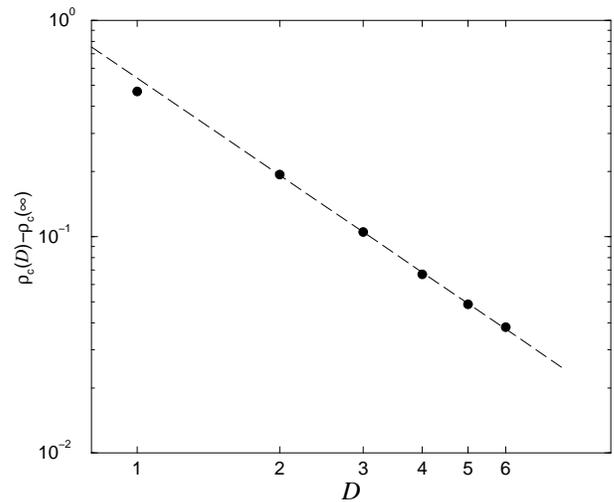}
  \caption{
    The critical density $\rho_{\text c}(D)$ as a function
    of the dimension~$D$.
    The critical density of the mean-field solution is denoted
    by $\rho_{\text c}(\infty)=1/2$.
    The dashed line corresponds to a power-law behavior 
    [Eq.\,(\protect\ref{eq:rho_c_d})]
    with an exponent $1.48$.
   }
  \label{fig:rho_c_d}
\end{figure}

\begin{figure}[b]
  \includegraphics[width=8.0cm,angle=0]{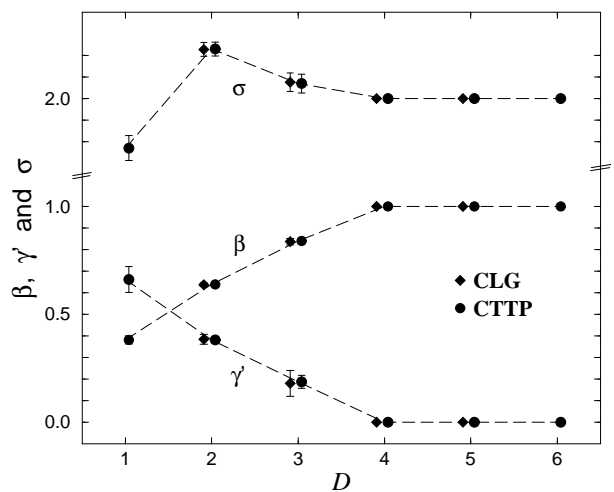}
  \caption{
    The critical exponents $\beta$, $\gamma^{\prime}$, and $\sigma$
    of the CTTP and CLG model (obtained from~\protect\cite{LUEB_19,LUEB_22})
    for various dimensions.
    The dashed lines are just to guide the eyes.
   }
  \label{fig:clg_cttp_exp}
\end{figure}

Finally we address the question how the 
critical densities depends on the dimension.
As can be seen from Table\,\ref{table:critical_indicees}
the critical density tends with increasing dimension
to the mean-field value $\rho_{\text c}=1/2$~\cite{LUEB_25}
that corresponds to an infinite dimension.
Our analysis reveals that the critical densities
approaches that mean-field value according to
\begin{equation}
\rho_{\text c}(D) \, - \, \frac{1}{2}
\; \sim \; D^{-\tau}  
\label{eq:rho_c_d}
\end{equation}
with $\tau=1.48\pm 0.05$ (see Fig.\,\ref{fig:rho_c_d}).
This behavior is different from that of CLG models 
on simple cubic lattices which is characterized by an 
exponent~$\tau=1$~\cite{LUEB_20}.

\section{Conclusions}
\label{sec:conc}

We investigated the steady state scaling behavior of the
CTTP model in various dimensions.
The order parameter exponent, 
the fluctuation exponent and the 
external field exponent are determined
and the corresponding values are listed in
Table\,\ref{table:critical_indicees}.
For $D=1$ and $D=2$ our results of the order parameter 
exponents differ from previous simulations
obtained from significantly smaller system sizes. 
Our values of the critical exponents $\beta$, 
$\gamma^{\prime}$, and $\sigma$ agree within the
error-bars with the corresponding exponents
of the CLG model (see Fig.\,\ref{fig:clg_cttp_exp}),
strongly supporting the conjecture~\cite{ROSSI_1}
that both models belong to the same 
universality class.

The picture is not so clear at the upper
critical dimension $D_{\text c}=4$.
Although the exponents are identical the 
logarithmic correction exponents 
of the CTTP and CLG model are different.
This result is rather surprising since the logarithmic
corrections exponents are a characteristic feature of the
whole universality class (see for instance~\cite{PFEUTY_1}).
We think that more than statistical uncertainties this
result is caused by systematic uncertainties of our
analysis.
In all cases we focused our attention to the leading
order of the scaling behavior.
Taking corrections to the leading order into 
account may result in comparable values of the
logarithmic correction exponents.
Further investigations are needed to clarify
this point.

\acknowledgments
I would like to thank A.~Hucht, H.-K.~Janssen,
and R.\,D.~Willmann for helpful discussions.
This work was financially supported by the 
Minerva Foundation (Max Planck Gesellschaft).

\begin{table}[b]
\caption{The critical density $\rho_{\text c}$ and 
the critical exponents $\beta$, $\sigma$, $\gamma^{\prime}$ and
$\gamma$ of the CTTP model for various dimensions~$D$.
The values of the susceptibility exponent~$\gamma$ are calculated
via Eq.\,(\protect\ref{eq:widom}).
The symbol $^{\ast}$ denotes logarithmic corrections
to the power-law behavior.}
\label{table:critical_indicees}
\begin{tabular}{llllll}
$\;D$       &  $\;\rho_{\text c}$	& $\;\beta$ &$\;\sigma$         & $\;\gamma^{\prime}$	& $\gamma$ \\  
\colrule \\
\tiny$\;1\;$ &  \tiny$\;0.96929\;$ & \tiny$\;0.382\pm0.019\;$ & \tiny$\;1.770\pm0.058\;$ & 
\tiny$\;0.662\pm0.071\;$  &\tiny$\;1.388\pm 0.063\;$ \\  
\tiny$\;2\;$ &  \tiny$\;0.69392\;$ & \tiny$\;0.639\pm0.009\;$ & \tiny$\;2.229\pm0.032\;$ & 
\tiny$\;0.381\pm0.013\;$  &\tiny$\;1.590\pm 0.033\;$ \\   
\tiny$\;3\;$ &  \tiny$\;0.60489\;$ & \tiny$\;0.840\pm0.012\;$ & \tiny$\;2.069\pm0.043\;$ & 
\tiny$\;0.187\pm0.030\;$  & \tiny$\;1.229\pm 0.045\;$ \\  
\tiny$\;4\;$ &  \tiny$\;0.56705\;$ & \tiny$\;1^{\ast}\;$      & \tiny$\;2^{\ast}\;$      & 
\tiny$\;0^{\ast}\;$       &\tiny$\;1^{\ast}\;$ \\  
\tiny$\;5\;$ &  \tiny$\;0.54864\;$ & \tiny$\;1\;$             & \tiny$\;2\;$             & 
\tiny$\;0\;$              & \tiny$\;1\;$ \\  
\tiny$\;6\;$ &  \tiny$\;0.53816\;$ & \tiny$\;1\;$             & \tiny$\;2\;$             & 
\tiny$\;0\;$              & $\;1\;$ 
\end{tabular}
\end{table}

\end{document}